\documentclass[conference]{IEEEtran}
\IEEEoverridecommandlockouts
\usepackage{cite}
\usepackage{amsmath,amssymb,amsfonts}
\usepackage{graphicx}
\usepackage{textcomp}
\usepackage{xcolor}
\usepackage[pscoord]{eso-pic}

\usepackage{algorithm}
\usepackage[noend]{algcompatible}

\algnewcommand{\IIF}[1]{\State\algorithmicif\ #1 \algorithmicthen}
\algnewcommand{\IELSE}{\State\algorithmicelse}
\algnewcommand{\EndIIF}{\unskip\ \algorithmicend\ \algorithmicif}

\begin{document}

\title{Acceleration of probabilistic reasoning through custom processor architecture
\thanks{This work has been supported by the EU ERC project Re-SENSE under grant agreement ERC-2016-STG-715037, and we acknowledge support form Intel.}
}

\author{\IEEEauthorblockN{Nimish Shah\IEEEauthorrefmark{1},
Laura I. Galindez Olascoaga\IEEEauthorrefmark{1}, Wannes Meert\IEEEauthorrefmark{2} and
Marian Verhelst\IEEEauthorrefmark{1}}
\IEEEauthorblockA{\IEEEauthorrefmark{1}MICAS, Department of Electrical Engineering, KU Leuven, Belgium\\
\IEEEauthorrefmark{2}DTAI, Department of Computer Science, KU Leuven, Belgium\\
Email: \{nimish.shah, laura.galindez, marian.verhelst\}@esat.kuleuven.be, wannes.meert@cs.kuleuven.be}}


\maketitle

\begin{abstract}
Probabilistic reasoning is an essential tool for robust decision-making systems because of its ability to explicitly handle real-world uncertainty, constraints and causal relations. Consequently, researchers are developing hybrid models by combining Deep Learning with probabilistic reasoning for safety-critical applications like self-driving vehicles, autonomous drones, etc. However, probabilistic reasoning kernels do not execute efficiently on CPUs or GPUs. This paper, therefore,
proposes a custom programmable processor to accelerate sum-product networks, an important probabilistic reasoning execution kernel. The processor has an optimized datapath architecture and memory hierarchy optimized for sum-product networks execution.
Experimental results show that the processor, while requiring fewer computational and memory units, achieves a 12x throughput benefit over the Nvidia Jetson TX2 embedded GPU platform.

\end{abstract}


\section{Introduction}

Deep learning (DL) has shown remarkable success in various computer vision and natural language processing tasks. However, it suffers from a serious drawback of not explicitly handling real-world uncertainty, constraints and causal relations, which are crucial for deploying robust decision-making systems in practice. To address this, researchers are developing hybrid models by combining DL with probabilistic reasoning techniques, eg. random sum-product networks \cite{peharz2019random}, probabilistic logic neural networks  \cite{qu2019probabilistic}, etc.


To ease the development of such hybrid models, several software frameworks like Pyro,
DeepProbLog,
etc.
have been developed. However, deploying these hybrid models in real applications 
raises demands of adequate
throughput, latency, energy consumption, etc. While several efficient hardware platforms have been developed for DL, probabilistic reasoning kernels are still implemented
on general-purpose CPUs
which is energy-inefficient. To address this, we propose a customized processor architecture that efficiently accelerates probabilistic reasoning workloads. Specifically, in this work:

\begin{itemize}
    \item GPU execution bottlenecks are identified by developing a highly-optimized CUDA implementation. 
    \item A programmable processor architecture is designed to alleviate these bottlenecks. 
    \item Experiments are performed to validate the performance improvement over a suite of benchmarks.
\end{itemize}

\begin{figure}[!t]
\centering
\includegraphics[trim={0cm 0cm 0cm 0cm} , clip, width=\columnwidth]{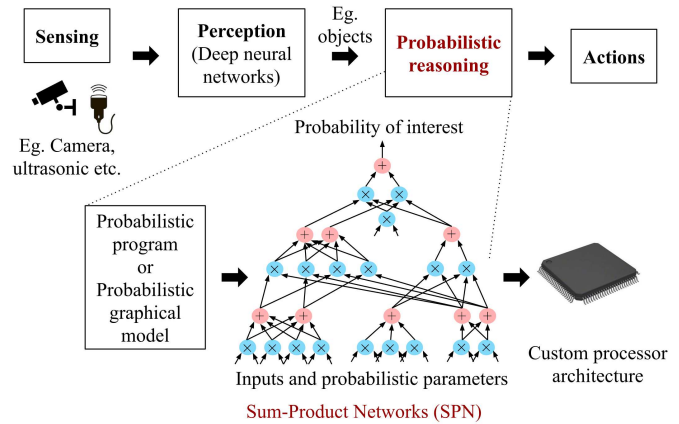}
\caption{An example of a hybrid system that uses Deep Learning for perception and probabilistic models for reasoning. Probabilistic models are usually converted to sum-product networks (SPN) for efficient inference.}%
\label{fig:overview}
\end{figure}
The paper is organized as follows: sec. \ref{sec:kernel} describes the compute kernel, while sec. \ref{sec:spn_on_cpu_gpu} discusses its performance on a CPU and GPU. The proposed processor architecture is described in sec. \ref{sec:processor}. Sections \ref{sec:results} and \ref{sec:discussion} evaluate the performance and discuss related work. Finally, sec. \ref{sec:conclusion} concludes the work.

\section{Probabilistic reasoning compute kernel} \label{sec:kernel}
As shown in fig. \ref{fig:overview}, probabilistic reasoning is usually performed with probabilistic program or a probabilistic graphical model.
These models can be converted 
to a tractable representation for efficient inference, called Sum-Product Network (SPN, also known as Arithmetic circuit) \cite{poon2011sum, chavira2008probabilistic}. Furthermore, SPNs can also be learned directly from data. 
SPNs are rooted directed acyclic graphs wherein the internal nodes are either sums or products, and the leaf nodes are probabilistic parameters or data inputs.

On a general-purpose processor, SPN can be implemented as a list of operations (alg. \ref{algo:spn_list}), or as a for loop over a vector (alg. \ref{algo:spn_for_loop}). This loop formulation already shows that, although SPNs offer parallelism, they inherently consist of scalar operations that consume prior intermediate results via irregular indirect memory accesses. This leads to inefficient parallel execution as will be demonstrated in the next section.

\begin{algorithm}
\caption{SPN as a list of operations}
\label{algo:spn_list}
\begin{algorithmic}[1]
\STATEx \textbf{Inputs}: \textbf{IN}: SPN leaf nodes as a vector
    \STATE r0 $=$ IN[0] $\times$ IN[1]
    \STATE r1 $=$ IN[2] $\times$ IN[3]
    \STATE r2 $=$ r0 $+$ r1
    \STATE ...    
    \STATE return rN
\end{algorithmic}%
\end{algorithm}




\begin{algorithm}
\caption{SPN as a for loop}
\label{algo:spn_for_loop}
\begin{algorithmic}[1]

\STATEx \textbf{Inputs}: \textbf{IN}: SPN leaf nodes as a vector of size \textbf{m}.
\textbf{O}: Vector of size \textbf{n} with binary indicators to select between sum or product for each arithmetic operation. \textbf{B},\textbf{C}: Vectors 
with pointers to first and second operand of each arithmetic operation.

    \STATE A $\gets$ [] // a vector of size m+n
    \STATE A[0...m-1] $\gets$ IN // initialize with inputs
    \FOR{i $=$ 0; i $<$ n; i$++$}
        \IF{O[i] $==$ SUM} 
            \STATE A[i+m] $=$ A[B[i]] $+$ A[C[i]]
        \ELSE
            \STATE A[i+m] $=$ A[B[i]] $\times$ A[C[i]]
        \ENDIF
    \ENDFOR
    \STATE return A[m+n-1]
\end{algorithmic}%
\end{algorithm}

\section{SPNs on CPU and GPU} \label{sec:spn_on_cpu_gpu}

To evaluate the performance of general-purpose platforms, SPNs are implemented on a CPU and a GPU.

\textbf{CPU}: An Intel core i5-7200 CPU is used to execute SPNs as lists of operations in C (alg. \ref{algo:spn_list}), giving the compiler maximal freedom to reorder and parallelize operations. The for-loop implementation (alg. \ref{algo:spn_for_loop}) is also evaluated, but it consistently performs slower than alg. \ref{algo:spn_list}

\textbf{GPU}: GPU offers parallelism in the form of a single instruction multiple thread (SIMT) mode. 
Multiple operations of SPNs can be executed in parallel by scheduling on multiple threads. However, the following two problems arise while doing so, due to the irregularity of SPN graphs.

\textbf{1) Inter-thread data transfers}: Inter-thread data transfers are needed when threads use data computed by other threads, which frequently happens when an SPN is scheduled across multiple threads. Such transfers can be allowed only after thread synchronization, as GPU does not guarantee lock-step execution of threads. To minimize these synchronization overheads in our experiments, SPNs are decomposed into groups of independent nodes (colored in fig. 2). As nodes in a group are not dependent on other nodes of the same group, they can be executed on any thread without synchronization. 
When executing a new group, threads still have to be synchronized with the CUDA primitive \texttt{\_\_syncthreads()}. Moreover, this inter-thread communication needs the data to be in either GPU \textit{shared} memory or \textit{global} memory. If data size fits, the shared memory is used for low-latency inter-thread transfers, otherwise the global memory is used with L1 caching enabled.

\textbf{2) Irregular memory accesses}: Data transfers from the shared memory become a bottleneck if multiple threads in a GPU warp access the same bank, resulting in bank conflicts.
These bank conflicts are minimized with a graph coloring-based bank allocation
that ensures that the threads in a warp access to different banks.

\begin{figure}[!t]
\centering
\includegraphics[trim={0cm 0cm 0cm 0cm} , clip, width=0.8\columnwidth]{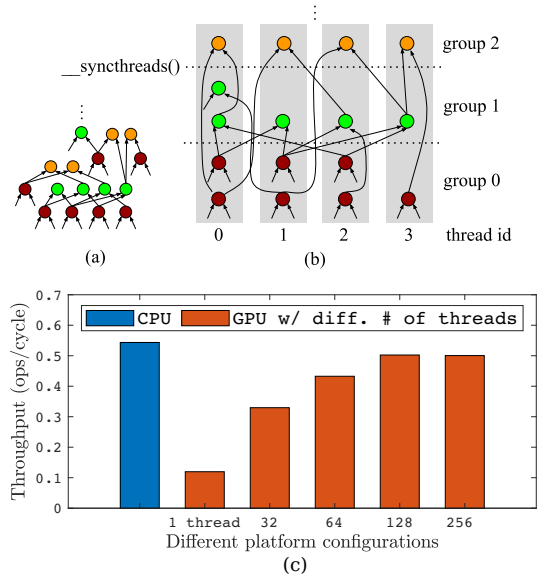}
\caption{Execution of an SPN on multiple GPU threads: (a) decompose SPN into '\textit{groups}' based on dependencies, and (b) execute on multiple threads and sync them before executing a new group. Throughput comparison of CPU and GPU in (c)}%
\label{fig:gpu_threads}%
\end{figure}%
The pseudocode of the developed CUDA implementation is shown in alg. \ref{algo:spn_cuda}. Several implementations are evaluated with varying number of threads in 1 thread block, on the Nvidia Jetson TX2 embedded GPU platform. An SPN trained on a benchmark in \cite{lowd2010learning} is used for performance measurement, with the resulting throughput shown in fig. \ref{fig:gpu_threads}(c). The GPU implementation utilizing a single thread expectedly performs worse than the CPU, as a GPU CUDA core is simpler than a superscalar CPU core. Use of 256 threads, however, only increases the throughput by a factor 4.1x, a sublinear scaling due to the following reasons: \begin{itemize}
    \item Overhead of thread synchronization \cite{letendre2013understanding}.
    \item Limited shared memory bandwidth. In Jetson TX2 GPU, shared memory has 32 banks among 128 CUDA cores. As all the threads read from (and write to) the shared memory, its bandwidth becomes a bottleneck.
    \item Thread divergence due to selection between sum and product operations leads to inactive threads in a warp.
\end{itemize}

An efficient processor for SPNs should strive for improved parallelization by avoiding these bottlenecks.


\begin{algorithm}
\caption{SPNcuda(IN, B, C, O)}
\label{algo:spn_cuda}
\begin{algorithmic}[1]
    \STATEx \textbf{kernel inputs}: Same as inputs of alg. \ref{algo:spn_for_loop}
    \STATE \texttt{int} i= blockDim.x*blockIdx.x + threadIdx.x //Thread ID
    \STATE \texttt{const int} t= TOTAL\_THREADS, x= length(IN)
    \STATE \texttt{\_\_shared\_\_} A[] // a vector in shared memory
    \STATE
    \FOR{j $=$ 0; j $<$ x; j$++$}
         \STATE A[i + j*t] $=$ IN[i + j*t] // copy inputs to shared memory
    \ENDFOR
    \STATE \texttt{\_\_syncthreads()}
    \STATE
    \STATE // compute first group of nodes
    \IF{O[i] == SUM}
        \STATE A[i + x*t] $=$ A[B[i]] $+$ A[C[i]]
    \ELSE
        \STATE A[i + x*t] $=$ A[B[i]] $\times$ A[C[i]]    
    \ENDIF
    \IF{O[i + t] == SUM}
        \STATE A[i + (x+1)*t] $=$ A[B[i + t]] $+$ A[C[i + t]]
    \ELSE
        \STATE A[i + (x+1)*t] $=$ A[B[i + t]] $\times$ A[C[i + t]]    
    \ENDIF
    \STATE ...
    \STATE \texttt{\_\_syncthreads()}
    \STATE // compute second group of nodes, and so on

\end{algorithmic}%
\end{algorithm}


\section{SPN processor architecture} \label{sec:processor}
\begin{figure}[!t]
\centering
\includegraphics[trim={0cm 0cm 0cm 0pt} , clip, width=\columnwidth]{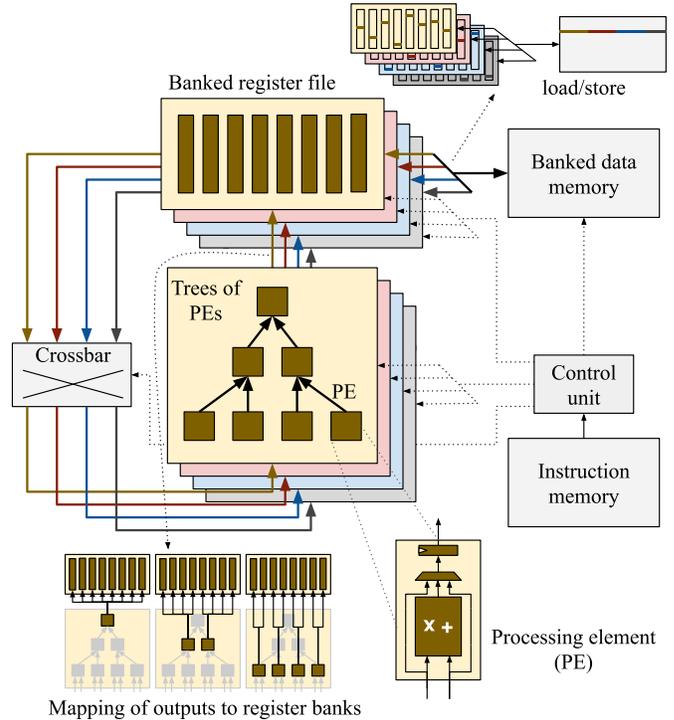}
\caption{The SPN processor architecture based on trees of processing elements (PE). A PE tree writes outputs to a private banked register file, while the inputs are serviced via a crossbar, allowing trees to access other register files. 
}%
\label{fig:processor}
\end{figure}

We propose a custom processor architecture (fig. \ref{fig:processor}) to alleviate the bottlenecks in SPN parallelization. It uses a light-weight method to exchange data among processing elements, and contains an appropriate memory hierarchy to handle irregular reads and writes efficiently. The important features are the following:

\textbf{Processing element (PE)}: The PE is a flexible arithmetic unit that can perform all the needed arithmetic operations (+,$\times$). It can additionally be configured to forward either of its inputs to output without performing any operation. The output of the PE is registered to allow adequate clock frequency.

\textbf{Trees of PEs}: The processor's datapath consists of \textit{trees} of PEs that enable local reuse of data, by avoiding frequent writebacks to the register file. 
Multiple trees allow execution of independent subgraphs of an SPN in parallel.

\textbf{Register file and register writes}: Each PE tree writes to a private banked register file. PEs at different levels of the tree can write to a different number of banks. As shown in fig. \ref{fig:processor} bottom left, PEs at the first level can write to 2 respective banks, the second level can write to 4 banks, and so on. 

\textbf{Crossbar and register reads}: The inputs of the PE trees are serviced by the register file via a crossbar, allowing PE trees to read any of the banks. Multiple inputs, however, cannot access the same bank in a given cycle. The crossbar is a combinational block with no internal memory.

\textbf{Data memory}: All the register files load/store data together in the form of a vector from a single address of the data memory, limiting irregular accesses to the register file, while requiring only vectorized transactions to the data memory.

\textbf{Programmability}: The architecture is made programmable with a custom VLIW instruction set, that can configure the trees and crossbar every clock cycle, copy data within register banks, and load/store from data memory.

\textbf{Compilation}: A custom compiler is developed to effectively schedule SPNs on the processor. The compiler directly takes as input the SPNs generated from tools like \cite{liang2017learning}. The compiler allocates register banks to intermediate results while trying to minimize the register read/write bank conflicts. This allocation has to happen in tandem with the placement of operation on the PE, since PEs cannot write to all register banks. Furthermore, the compiler also reorders operations to minimize read-after-write hazards due to pipelining in PE trees. Finally, the intermediate data from the register files are spilled carefully to the data memory to minimize load-stores. In the end, the compiler generates a list of the custom VLIW instructions that can be executed directly on the processor.


In summary, the architecture allows local reuse of intermediate results by using \textit{trees} of PEs, enables flexible movement of data with the crossbar, and efficiently supports irregular data fetches by using banked register files. The custom compiler allows to effectively use these architectural features. 

\section{Experimental results} \label{sec:results}
We experimentally benchmark the throughput of the proposed processor architecture against optimized CPU and GPU implementations, with SPNs trained on a suite of standard benchmarks \cite{lowd2010learning,Dua:2019} using the algorithm in \cite{liang2017learning}. 
The CPU and GPU implementations are as described in sec. \ref{sec:spn_on_cpu_gpu}. For a fair comparison, the performance of the proposed processor is evaluated for configurations that consume less resources than the GPU (summarized in Table \ref{tab:platforms}): 32 banks in the register file, same as the amount of GPU shared memory banks, with 64 registers in each bank. 

\begin{table}[!b]
\caption{Compute and memory details of different processing platforms}
\label{tab:platforms}
\centering
\begin{tabular}{|c|c|c|c|}
\hline
\textbf{Platform} & \textbf{\begin{tabular}[c]{@{}c@{}}Compute\\ units\end{tabular}} & \textbf{\begin{tabular}[c]{@{}c@{}}Immediate\\ memory size\end{tabular}} & \textbf{\begin{tabular}[c]{@{}c@{}}memory\\ banks\end{tabular}} \\ \hline
CPU & \begin{tabular}[c]{@{}c@{}}2 arith. units in\\a superscalar core \end{tabular} & \begin{tabular}[c]{@{}c@{}}168 80b registers\\ 32 KB L1 cache\end{tabular} & 16 \\ \hline
GPU & \begin{tabular}[c]{@{}c@{}}128 \\ CUDA cores\end{tabular} & \begin{tabular}[c]{@{}c@{}}64K 32b registers\\ 64 KB shared mem.\end{tabular} & 32 \\ \hline
\begin{tabular}[c]{@{}c@{}}Ours\\ \textit{(Pvect)}\end{tabular} & 16 PEs & \begin{tabular}[c]{@{}c@{}}2K 32b registers \\ 64 KB data mem.\end{tabular} & 32 \\ \hline
\begin{tabular}[c]{@{}c@{}}Ours\\ \textit{(Ptree)}\end{tabular} & 30 PEs & \begin{tabular}[c]{@{}c@{}}2K 32b registers \\ 64 KB data mem. \end{tabular} & 32 \\ \hline
\end{tabular}

\end{table}

\begin{figure}[!b]
\centering
\includegraphics[trim={0cm 0cm 0cm 0.5cm} , clip, width=\columnwidth]{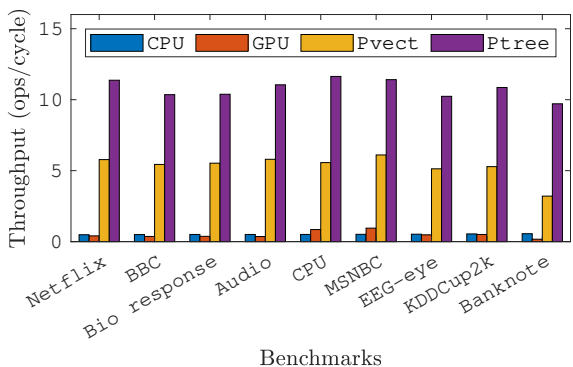}
\caption{Throughput comparison of different platforms on several benchmarks. 
}%
\label{fig:throughput}
\end{figure}

Two configurations of the custom processor with different arrangements of PEs are evaluated to highlight the impact of the tree arrangement. The \textit{\textbf{Ptree}} configuration uses 2 trees each with 4 levels of PEs (30 PEs in total).
The \textit{\textbf{Pvect}} configuration removes the trees and uses only the lowest level PEs (16 PEs in total). Note that both configurations are identical in every aspect (crossbar, register file etc.), except for the arrangement and the number of PEs. 

Throughput for the CPU and GPU is measured by averaging the runtime for 100k SPN executions, resulting in the peak performances of 0.55 and 0.95 effective operations/cycle respectively. For accurate throughput measurement of our processor, a cycle-accurate model is developed in the MyHDL framework \cite{myhdl}. The resulting performance is compared in fig. \ref{fig:throughput} in terms of operations/cycle. The throughput of \textit{Ptree} is at least 12x higher than the CPU and GPU, with a peak performance of 11.6 operations/cycle. Moreover, \textit{Ptree} performs 2x better than \textit{Pvect}, confirming the benefit of the tree arrangement of the PEs. 

\section{Related work} \label{sec:discussion}
In \cite{sommer2018automatic}, authors implemented SPNs on an Nvidia 1080Ti discrete GPU using Tensorflow 
and observed much lower throughput than a CPU. However, Tensorflow is ill-suited for SPNs as it launches a new GPU kernel for every operation in SPN, incurring prohibitively high overhead. We show that a custom CUDA-based implementation can achieve throughput similar to a CPU, even on an embedded GPU platform. 

In recent years, a few works have proposed custom hardware for SPNs acceleration. Work in \cite{shah2019problp, sommer2018automatic,zermani2015bayesian} uses fully-parallel spatial architectures that implement every operation in SPNs as a unique hardware operator. This is typically intended for FPGA implementation. These fully unrolled implementations, however, are not scalable to the large SPNs required for real applications. In \cite{schumann2015towards}, authors proposed a processor with multiple processing elements that use a common data bus to share intermediate results,
making the bus a severe bottleneck. The processor in \cite{lin2010high}, 
proposed for a different probabilistic reasoning workload, has similarities with our work. However, it has unnecessary hardware redundancy (multiple crossbars) while having stringent requirements for the graph structure (only polytrees), making it unsuitable for SPNs. Our work retains it's advantages while being more general.

\section{Conclusion} \label{sec:conclusion}
An efficient processor architecture is proposed to accelerate Sum-product networks, an important compute kernel for probabilistic reasoning. The processor contains hardware features suitable for the irregular graph processing of SPNs, such as trees of processing elements for data reuse, a crossbar for flexible sharing of data, and independent-addressable register banks for irregular data reads and writes. 
The performance of the processor is compared with CPU and an highly-optimized GPU CUDA implementation, showing 12x higher throughput while using fewer computational and memory units. 


\bibliographystyle{IEEEtranS}
\bibliography{ref_old}

\end{document}